\documentclass[reprint, superscriptaddress, amsmath, amssymb, aps, prl, floatfix]{revtex4-2}

\usepackage{graphicx}
\usepackage{dcolumn}
\usepackage{bm}
\usepackage{commath}
\usepackage{physics}

\begin{document}

\preprint{APS/123-QED}

\title{High Magnetoresistance Ratio on hBN Boron-Vacancy/Graphene Magnetic Tunnel Junction}

\author{Halimah Harfah} 
\email{Corresponding author: harfah.h@opt.mp.es.osaka-u.ac.jp}
\thanks{These authors contributed equally to this work}
\affiliation{Graduate School of Engineering Science, Osaka University, 1-3 Machikaneyama-Cho, Toyonaka, Osaka 560-0043, Japan }

\author{Yusuf Wicaksono}
\thanks{These authors contributed equally to this work}
\affiliation{Computational Condensed Matter Physics Laboratory, RIKEN, 2-1 Hirosawa, Wako, Saitama 351-0198, Japan}

\author{Gagus Ketut Sunnardianto}
\affiliation{
Research Center for Quantum Physics, National Research and Innovation Agency (BRIN), Tangerang Selatan, Banten, 15314, Indonesia
}
\affiliation{Research Collaboration Center for Quantum Technology 2.0, Bandung 40132, Indonesia}

\author{Muhammad Aziz Majidi}
\affiliation{
 Department of Physics, Faculty of Mathematics and Natural Science, Universitas Indonesia, Kampus UI Depok, Depok 16424, Indonesia.
}

\author{Koichi Kusakabe}%
\affiliation{
 School of Science, Graduate School of Science, University of Hyogo 3-2-1 Kouto, Kamigori-cho, Ako-gun, 678-1297, Hyogo, Japan.
}

\date{\today}

\begin{abstract}
We presents a new strategy to create a van der Waals-based magnetic tunnel junction (MTJ) that consists of a three-atom layer thickness of graphene (Gr) sandwiched with hexagonal boron nitride (hBN) by introducing a monoatomic Boron vacancy in both hBN layers. The magnetic properties and electronic structure of the system were investigated using density functional theory (DFT), while the transmission probability of the MTJ was investigated using the Landauer--B\"uttiker formalism within the non-equilibrium Green function method. The Stoner gap was found to be created between the spin-majority channel and the spin-minority channel on LDOS of the hBN monoatomic boron-vacancy (V$_B$) near the vicinity of Fermi energy, creating a possible control of the spin valve by considering two different magnetic allignment of hBN(V$_B$) layers, anti-parallel and parallel configuration. The results of the transmission probability calculation showed a high electron transmission in the parallel configuration of the hBN(V$_B$) layers and a low transmission when the antiparallel configuration was considered. A high TMR ratio of approximately 400\% was observed when comparing the antiparallel and parallel configuration of hBN(V$_B$) layers in the hBN (V$_B$)/Gr/hBN(V$_B$), giving the highest TMR for the thinnest MTJ system.
\end{abstract}

\maketitle

\section{Introduction}
Magnetic tunnel junctions (MTJs) have been extensively studied for their potential applications in spintronics, such as logic devices, hard drive magnetic read heads, and magnetic sensors \cite{Chappert:2007,Dery:2007,Zhu:2006,Childress:2005,Chen:2010,Iqbal:2018-review}. The tunneling magnetoresistance (TMR) ratio is a key factor in improving MTJs, and MgO is the most common tunnel barrier used. CoFeB/MgO/CoFeB MTJs have achieved the highest TMR of 1100\% at 4.2 K \cite{ikeda:2010}. However, when the barrier thickness is reduced to downscale the device, the TMR can be reduced to 55\% due to the presence of uncontrollable defects in the MgO tunnel barrier \cite{Reiss2013:thin-film_MgO, Wang2011}.

Researchers have recently turned their attention to a 2D material-based magnetic junction for novel spintronics applications. 2D materials such as graphene or hBN have been used as a spin valve in the current-in-plane (CIP) scheme \cite{yusuf1,yusuf2,yusuf3,mori} by controlling its unique topological nature or have been used as non-magnetic spacers in current-perpendicular-to-plane (CPP) MTJs. While experimental data supporting the effectiveness of the former have not yet been confirmed, the latter has shown relatively low performance in the TMR ratio \cite{Piquemal_Banci:2017,Exfolation1,Asshoff:2017,Iqbal:2015-JMCC,Iqbal:2013-NanoRes,Chen:2013,Li:2014,LiWan:2014,Mandal:2012,Martin:2015,Entani:2016}. Previous studies show that the interface between the ferromagnetic electrode (FME) and 2D materials plays an important role in transmission through the tunnel barrier \cite{Lu2021,Harfah2020,Harfah2022}. When a conventional FME was created, that is, an electrode of a transition metal such as Fe, Ni, or Co, a chemisorption was created that resulted in a strong $pd$ hybridization between 2D materials and FME \cite{Lu2021,Harfah2020,Harfah2022}. This hybridization affects the transmission of electrons, especially when 2D polar materials, such as hBN, are used as a tunnel barrier. Strong $pd$ hybridization at the interface leads to buckling of the hBN and creates a dipole moment at the interface. This dipole moment leads to the shift of Fermi energy from the peak of transmission through the $d$-orbital of FME, creating a small TMR ratio \cite{Lu2021,Harfah2020,Harfah2022}. 

Several strategies have been proposed to avoid a strong $pd$-hybridization between FME and hBN, such as the introduction of transition metal FME alloys with inert metals such as Pt creating Co$_3$Pt alloy \cite{Lu2021} or using MXene-based FME\cite{Wang2018,Wen2023,Zhou2020}. The occurrence of Pt in the alloy allows physisorption of hBN on the FME instead of chemisorption\cite{Lu2021}, creating a high transmission peak in parallel configuration (PC) at Fermi energy. Similarly, MXene-based FME, which usually consists of a rare-earth heavy-element metal such as Fe$_3$GeTe$_2$, creates a van der Walls interaction with hBN due to the relatively large atomic size and a more diffuse electron of heavy-elements atom (Te atom) cloud compared to lighter elements\cite{Wang2018}. However, creating and maintaining a long-range order of single-phase structure on the surface of Co$_3$Pt alloy or Fe$_3$GeTe$_2$ is rather challenging, especially at room temperature and high production cost \cite{Zhou2020,Front2021,Liu2022}. 

On the other hand, the magnetic properties of hBN Boron-vacancy (V$_B$) has been a subject of significant research interest for the purpose of quantum computing applications \cite{hBN-VB1,hBN-VB2,hBN-VB3}. Recent experiments recorded optically detected magnetic resonance (ODMR) signals at room temperature for color centers in hBN(V$_B$) \cite{hBN-VB1}. In addition, distinct resonances are observed in the room-temperature ODMR spectrum even without an external magnetic field \cite{hBN-VB2}. This observation further supports the presence of the magnetic moment associated with the V$_B$ defect at room temperature. A first-principles study also supports the experimental results of the ODMR signals \cite{hBN-VB3}, confirming a possible magnetic moment on hBN(V$_B$) in room-temperature. Although relatively low photoluminescence and contrast of ODMR were found \cite{hBN-VB4}, which means that further research is necessary to enhance the ODMR sensitivity in hBN spin defects for the quantum computing application, the apperance of the magnetic moment on the hBN(V$_B$) becomes a possible key point to create a successful design of the thinnest MTJ using 2D materials.

In this work, we present a new strategy, employing light-element-based materials, to create a van der Waals-based MTJ with only a three-atom-layer thickness. Here, we proposed a three layer of hBN(V$_B$)/Graphene(Gr)/hBN(V$_B$) sandwiched structure. hBN (V$_B$) is used as the ferromagnetic layer that filters the transmission of electrons from two non-magnetic electrodes. The Gr layer is used between hBN (V$_B$) to keep transmission not too small due to the insulating nature of hBN \cite{Harfah2022}. The investigation of the magnetic properties and electronic structure of the system was performed using density functional theory (DFT), while the transmission probability of hBN(V$_B$)/Gr/hBN(V$_B$) MTJ was investigated using the Landauer--B\"uttiker (LB) formalism within the non--equilibrium Green function (NEGF) method. Our calculation results observed a high TMR ratio of about 400\% in the hBN(V$_B$)/Gr/hBN(V$_B$) MTJ with Cu metal used as electrode. This results show the highest TMR ratio found in the thinnest MTJ system, which consists only of a three-atomic layer thickness. 

\section{Computational Method}

\begin{figure}[tb]
\includegraphics[width=\columnwidth]{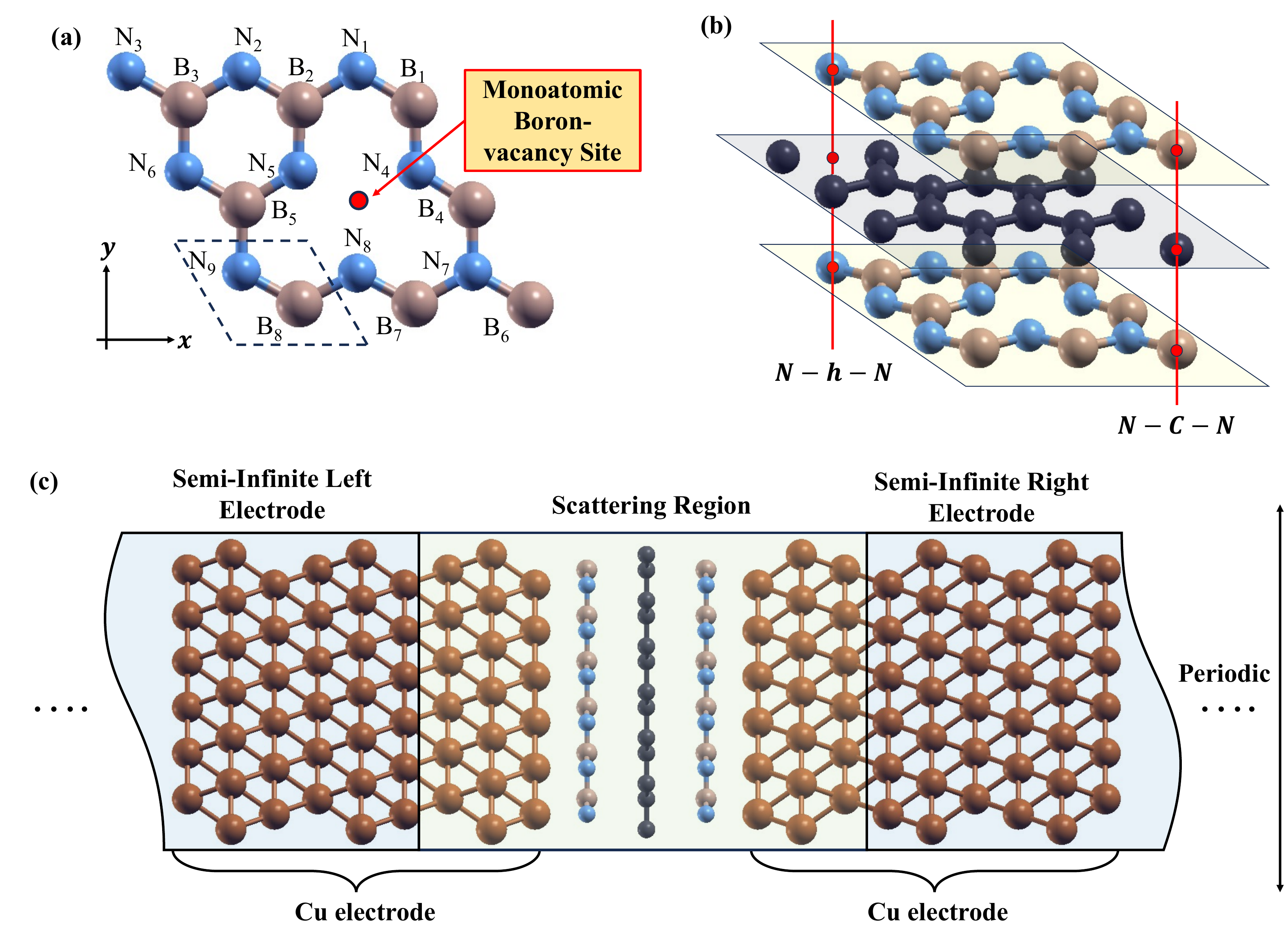}
\caption{\label{fig:figure1} (a) The crystal structure of hBN with Boron-vacancy. a $3 \times 3$ unit cell was used in the calculations which consists of nine atoms of Nitrogen and eight atoms of Boron. Blue (pink) color ball represents N (B) atoms. (b) hBN(V$_B$)/Gr/hBN(V$_B$) MTJ system where N atoms of hBN right on top and below hollow-site of Gr ($N-h-N$) and B atoms right on top and below of C atoms of Gr ($B-C-B$). The black color ball represent C atoms of graphene. (c) The setup to calculate the tunneling transmission probability of hBN(V$_B$)/Gr/hBN(V$_B$) MTJ where Cu is used as electrode. The scattering region consists of 3 layer Cu/hBN(V$_B$)/Gr/hBN(V$_B$)/3 layer Cu, while both left and right electrode consists of 6 layer of Cu. }
\end{figure}

In this theoretical work, a supercell of $3 \times 3$ hBN layer with monoatomic atom B vacancy was considered, as shown in Figure \ref{fig:figure1}(a). In the MTJ system, a Gr layer is sandwiched between hBN(V$_B$) layers, as shown in Figure \ref{fig:figure1}(b). The most stable configuration of hBN/Gr/hBN was observed where the B atoms of the hBN were on top and below the C atoms of graphene, while the N atoms were right on top and below the hollow site of graphene\cite{Harfah2022,yusuf3,quhe_2012}. Since the Boron monovacancy in hBN does not change much in the van der Walls interaction between hBN and graphene in $3 \times 3$ supercell, this stable stacking configuration was also applied on the hBN(V$_B$)/Gr/hBN(V$_B$) MTJ system, as shown in Figure \ref{fig:figure1}(b). In the case of hBN(V$_B$)/Gr/hBN(V$_B$) MTJ investigation, both the antiparallel configuration (APC) and the parallel configuration (PC) were considered for the upper and lower hBN(V$_B$). In the case of APC, the magnetic moment orientation of the lower (upper) hBN(V$_B$) was fixed in the upward (downward) direction. On the other hand, both upper and lower hBN(V$_B$) magnetic moment orientations were fixed upward when the PC state was considered.

The SIESTA package \cite{siesta1,siesta2} was employed to calculate structural equilibrium, magnetic properties, spin-charge density mapping, and local density of states (LDOS) using spin-polarized DFT. The Troullier--Martins \cite{tmpseudo} pseudopotential and a revised Perdew--Burke--Ernzerhof functional for a densely packed solid surface (i.e., the PBESol functional) \cite{PBEsol} were used to describe the electron--ion interaction within the generalized gradient approximation (GGA). A basis set with double-zeta and polarization was utilized \cite{dzp1,dzp2,dzp3}. The atomic positions were relaxed with a 0.001 eV/\AA~ force tolerance, and a $36 \times 36 \times 1$ Monkhorst--Pack k-mesh was employed for the calculations. An 800 Ry mesh cut-off was also used. Additionally, van der Waals interactions between hBN and graphene were taken into account by applying a Grimme-type dispersion potential \cite{grimme}.

The tunneling transmission probability will be calculated using the LB formalism within the NEGF method. First, the tunneling transmission probability will be done on hBN(V$_B$) to understand the capability of hBN(V$_B$) layer filtered the electron spin transmission through the CPP scheme. Subsequently, the tunneling transmission probability calculation was done on hBN(V$_B$)/Gr/hBN(V$_B$) MTJ. For the purpose of calculating the tunneling transmission probability, a Cu electrode is used in the calculation, which is a non-magnetic electrode, as shown in Figure \ref{fig:figure1}(c). The spin-dependent current was calculated using the LB equation given by:

\begin{equation}\label{eq:LB_method}
    I^{\uparrow(\downarrow)} = \frac{e}{h} \int_{\min \infty}^\infty T^{\uparrow(\downarrow)}(E)\Big[f_L(E,\mu) - f_R(E,\mu)\Big]dE
\end{equation}

\noindent where $f_L(E,\mu)\big(f_R(E,\mu)\big)$ is the right (left) moving electron injected from the left (right) lead in the form of the Fermi–Dirac function. $\mu_L\big(\mu_R\big)$ denotes the chemical potentials of the left (right) electrodes. Since the zero bias voltage was considered, thus $\mu_L = \mu_R = E_F$ . In addition, the ballistic transmission $T$ as a function of energy $E$ is described with respect to the Green function form as

\begin{equation}\label{eq:T-matrix_NEGF}
    T^{\uparrow(\downarrow)}(E) = \Tr{\Big[\Gamma_LG^R\Gamma_RG^A}
\end{equation}

\noindent where $\Gamma_L(\Gamma_R)$is the coupling matrix of the left (right) electrode, $G^R(G^A)$ is the retarded (advanced) Green functions of the central region.

Finally, in the hBN(V$_B$)/Gr/hBN(V$_B$) MTJ system, the MR ratio can be calculated by including the difference between the current in the APC and PC states and then dividing it by the current in the APC state as follows:

\begin{equation}
    \textrm{TMR ratio} = \frac{I_{PC} - I_{APC}}{I_{APC}}\times 100\%.
    \label{eq:MR_ratio}
\end{equation}

\section{Results and Discussion}
\subsection{Magnetic, electronics, and transport properties of hBN(V$_B$) as spin-filter}

\begin{figure}[tb]
\includegraphics[width=\columnwidth]{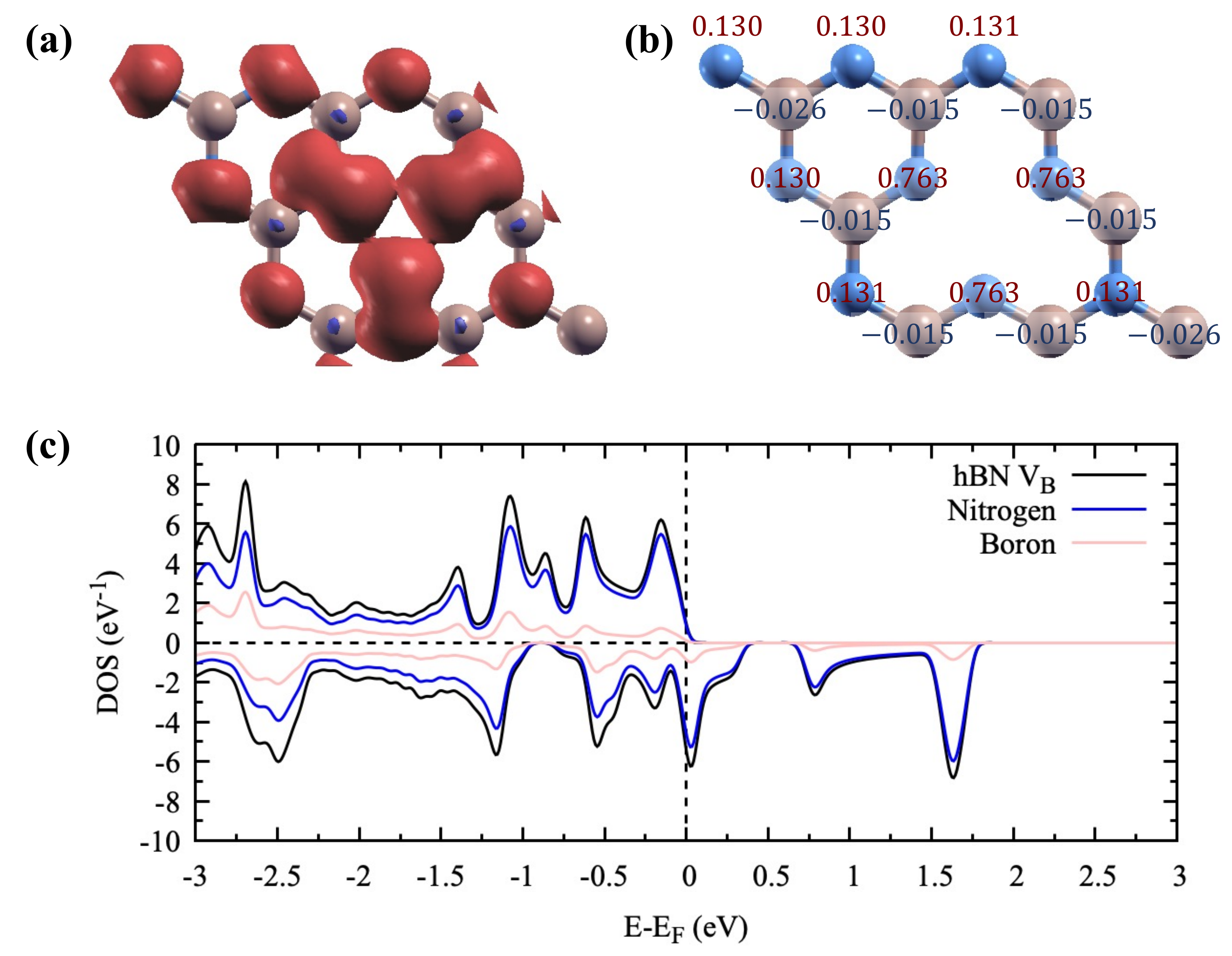}
\caption{\label{fig:figure2} (a) The hBN $3 \times 3$ V$_B$ spin charge density mapping (SCDM) (red (blue) color represent spin-up (spin-down) electorn density), (b) magnetic moment on each atoms (in $\mu_B$ unit; negative value corresponds to spin-down and positive value corresponds to spin-up), and (c) local density of states (LDOS). The positive (negative) value of DOS represents spin majority (minority) channel.}
\end{figure}

The monoatomic Boron vacancy in the hBN layer was expected to give magnetization in the hBN due to the unpaired electron of the N atoms at the vacancy site due to the absence of a $\sigma$-bond with the B atom. We performed spin charge density mapping (SCDM) to our proposed $3 \times 3$ hBN with monoatomic Boron vacancy to understand the distribution of spin-up and spin-down electron density. Figure \ref{fig:figure2}(a) shows that the spin-up electron density dominates the hBN(V$_B$), due to our initial configuration of the system having a spin-up magnetization direction. From SCDM it shows that the spin-up electron density was found to be strong near the vacancy. However, the N atoms relatively far from the vacancy retain the strong spin-up electron density mapping, but the amplitude was reduced. Meanwhile, the B atoms have spin-down electron density, which is opposite to that of the N atoms. A small spin-down electron density was expected in the B atoms, since all valence electrons of the B atoms were used to create a $\sigma$-bond with the N atoms. 

\begin{figure*}[tb]
\includegraphics[width=\textwidth]{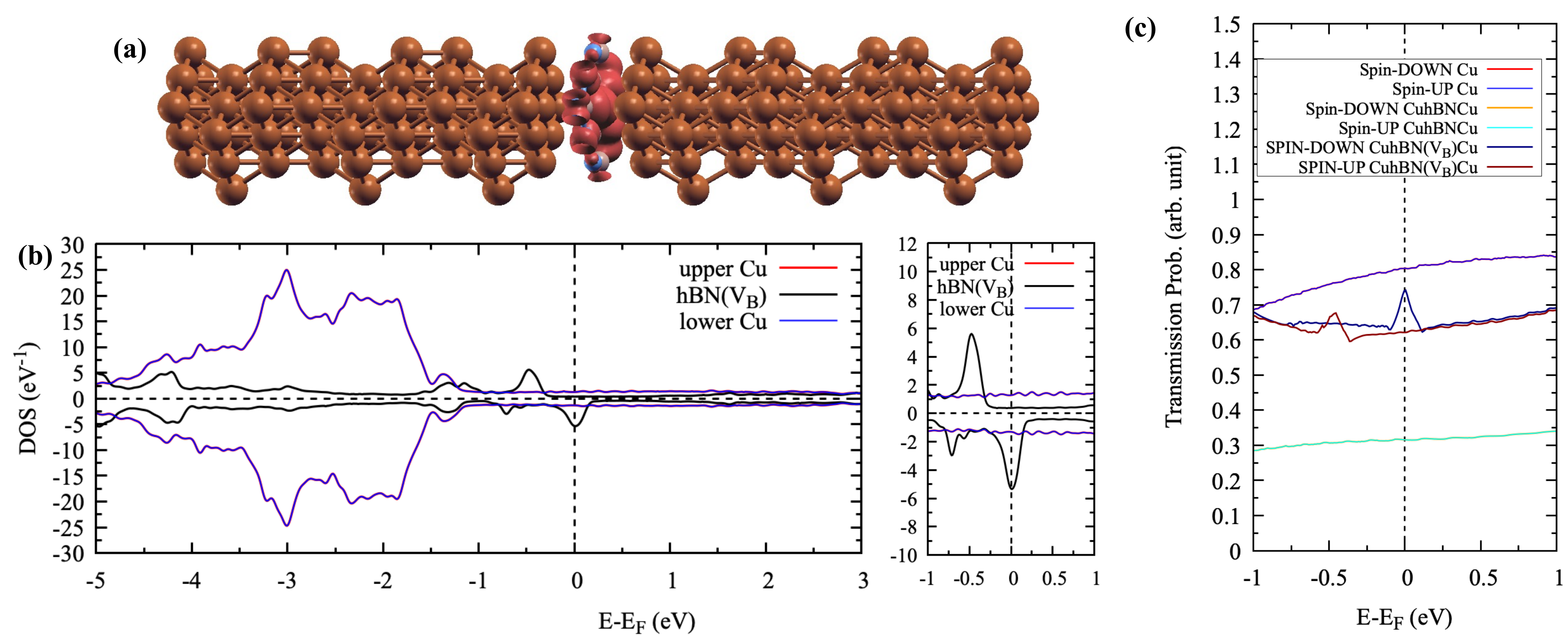}
\caption{\label{fig:figure3} (a) SCDM of $3 \times 3$ Cu/hBN(V$_B$)/Cu (red (blue) color represents spin-up (spin-down) electorn density). (b) LDOS of $3 \times 3$ Cu/hBN(V$_B$)/Cu. The upper (lower)-Cu corresponds to the Cu layer right on top (below) of hBN(V$_B$). The positive (negative) value of DOS represents the spin majority (minority) channel. (c) Tunneling transmission probability of Cu/hBN(V$_B$)/Cu with Cu/hBN/Cu and pristine Cu transmission probability as a reference. }
\end{figure*}

Figure \ref{fig:figure2}(b) shows the magnetic moment of each atom in hBN(V$_B$). The magnetic moment of N atoms near the vacancy implies that the spin-up electron accumulated near the vacancy from the unpaired electron of N atoms has a magnetic moment around 1.5 $\mu_B$. The remaining magnetic moment in N atoms at the vacancy, which around $\approx 0.263$ comes from the unpaired electron at the $p_z$-orbital of N atoms. This result suggests that a ferromagnetic order was dominant between the magnetic moment near the vacancy and N atoms. The reduced value of the magnetic moment on N atoms, which are relatively far from vacancy, indicated a short range of magnetic exchange was created. Note that when a bigger supercell was considered, local ferromagnetism is expected in the hBN(V$_B$) layer, instead of creating a full ferromagnetic layer.

The LDOS of the $3 \times 3$ hBN(V$_B$) layer was shown in Figure \ref{fig:figure2}(c). The DOS of state near Fermi energy shows that the Stoner gap was created between the spin majority channel and the spin minority channel. Since the magnetic moment direction was set to spin-up direction, the spin majority channel has shifted to a lower energy, creating a higher occupied state compared to the spin minority channel. In the spin-majority channel, the huge band gap was observed from the Fermi energy to the higher energy, which represents a typical huge band gap of hBN and represents its insulating nature. However, in the spin minority channel, additional states were shown to be created from $0.6$ eV to $1.8$ eV. This localized state in the middle of the hBN insulator gap corresponds to the localized state of the vacancy. Since the magnetic moment was set into a spin-up direction, the localized state at the insulator gap of hBN in the spin minority channel is unoccupied (at an energy higher than the Fermi energy). Furthermore, it is also shown that this localized state was mainly coming from the N atoms by looking at the LDOS of the N atoms in the hBN (V$_B$) layer. 

The Stoner gap found in the DOS of hBN becomes a potential aspect to be used for electron spin filter when a non-magnetic electrode is connected, creating a CPP scheme of electron transport. At the Fermi energy, it was shown that a high-spin filter is possible where the spin-majority channel has lower density while, oppositely, the spin-minority channel has high density. To explore this possibility, a tunneling transmission probability was carried out on the $3 \times 3$ hBN(V$_B$) by connecting it to the Cu electrode on the upper and lower part, as shown in Figure \ref{fig:figure3}(a). When the hBN layer comes into contact with the Cu layer, a physisorption between the Cu/hBN interface is expected \cite{Lu2021}. This physisorption at the Cu/hBN interface was also applied at the Cu/hBN(V$_B$) where the interlayer distance between Cu at the interface and hBN is around $2.51$\AA. 

When Cu/hBN has a physisorption interface, although no hybridization was created between Cu and hbN, a proximity effect was observed on the hBN layer, changing the electronic structure from an insulator to a metal (see Section III Supplementary Materials). This proximity effect was also found in physisorption at the interface of Cu/hBN(V$_B$). The LDOS of hBN in Cu/hBN(V$_B$)/Cu shows a small but non-zero density of states near the Fermi energy, as shown in Figure \ref{fig:figure3}(b), indicating a metal electronic structure. However, although the proximity effect changes the hBN(V$_B$) slightly metalic, the induced magnetic moment on hBN(V$_B$) still kept as shown in Figure \ref{fig:figure3}(a) where there is a clear spin-up charge density on hBN(V$_B$). This is also confirmed by the Stoner gap created on the hBN(V$_B$) LDOS, especially near the Fermi energy, as shown on the right side of Figure \ref{fig:figure3}(b). It is also shown that the DOS from the vacancy site still survives, creating two DOS peak on the hBN (V$_B$) LDOS at Fermi energy ($E-E_F=-0.5$ eV) for the spin minority (majority) channel. It should be noted in Figure \ref{fig:figure3}(a) that the induced magnetization on hBN due to the vacancy does not give rise to an induced magnetic moment on the Cu layer even at the interface. Figure \ref{fig:figure3}(b) also shows that no Stoner gap was created in the Cu layers (both at the top and below hBN(V$_B$)), implying no magnetic moment was induced on the Cu layer at the interface.

The Stoner gap in the hBN(V$_B$) LDOS at the interface is expected to give a spin-filter effect when current flows from the left electrode to the right electrode (vice versa). The calculation of the tunneling transmission probability on Cu/hBN(V$_B$)/Cu shown in Figure \ref{fig:figure3}(c) suggests a spin filter effect at the Fermi energy. This spin-filter was mediated by the localized state of the vacancy site, since the peak transmission of the spin-down (spin-up) electron corresponds to the LDOS of the vacancy site in the spin minority (majority) channel. Furthermore, by comparing the transmission probability of pristine Cu, Cu/hBN/Cu, and Cu/hBN(V$_B$)/Cu, the localized state of the vacancy site transmit electron is more efficient than that of pristine hBN, shown by the slight reduction in the transmission probability of the electron compared to that of pristine Cu. The efficient spin-filter mediated by the localized state of hBN(V$_B$) gives a prospective spin-valve device when a sandwiched structure of hBN(V$_B$)/Gr/hBN(V$_B$) is considered, which is discussed in the next section.

\subsection{The high tunneling magnetoresistance ratio on hBN(V$_B$)/Gr/hBN(V$_B$) MTJ}

\begin{figure}[tb]
\includegraphics[width=\columnwidth]{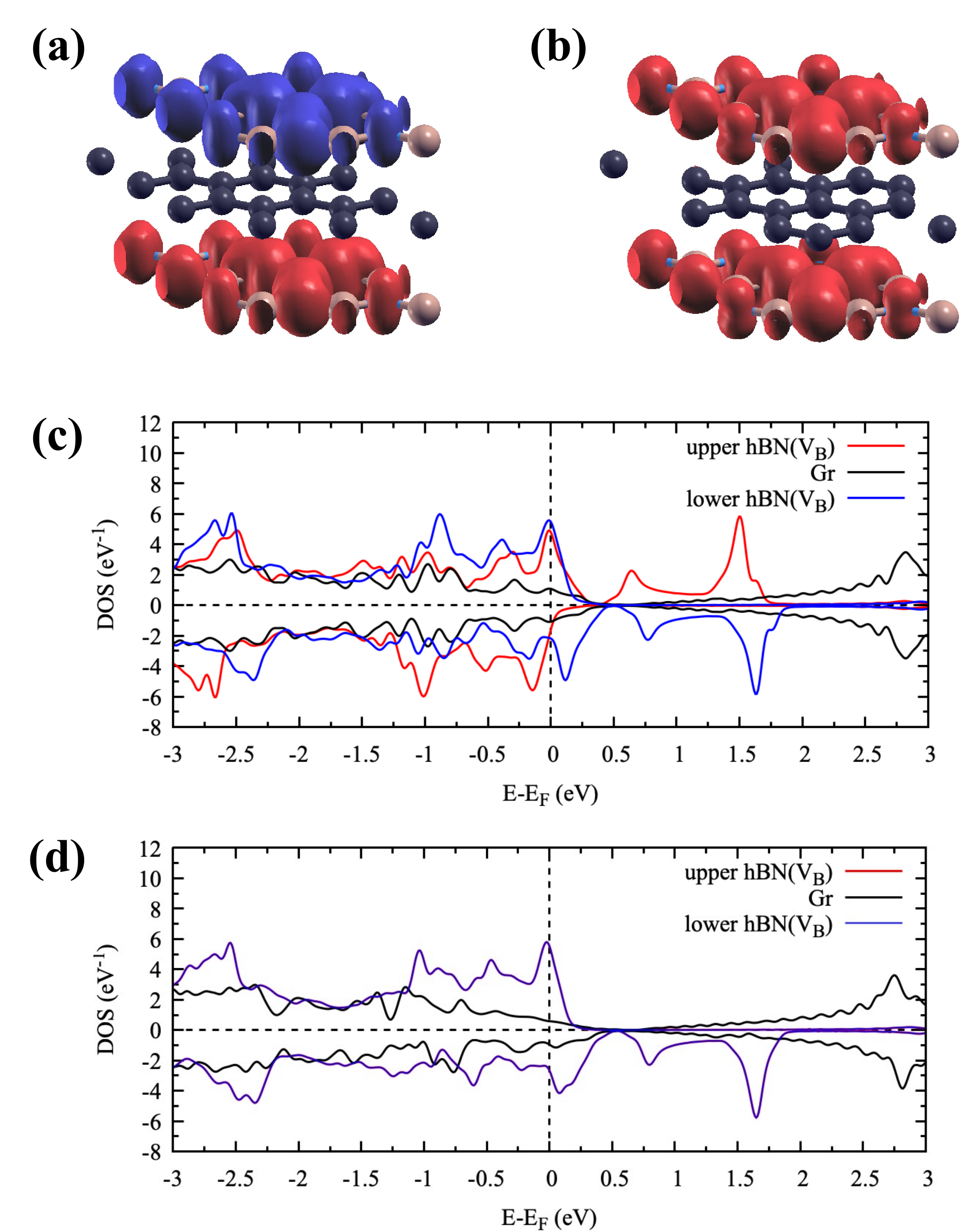}
\caption{\label{fig:figure4} The SCDM of $3 \times 3$ hBN(V$_B$)/Gr/hBN(V$_B$) for (a) APC and (b) PC states (red (blue) color represent spin-up (spin-down) electorn density). LDOS of $3 \times 3$ hBN(V$_B$)/Gr/hBN(V$_B$) for (c) APC and (d) PC states. The positive (negative) value of DOS represents spin majority (minority) channel.}
\end{figure}

To create an MTJ system based on hBN(V$_B$), hBN(V$_B$)/Gr/hBN(V$_B$) was considered. Two magnetic configurations of antiparallel configuration (APC) and parallel configuration (PC) of the upper and lower hBN(V$_B$) layers were considered. The SCDM of both magnetic configurations is shown in Figures \ref{fig:figure4} (a) and (b). The lowest energetically magnetic configuration is the PC state, with the total energy difference between the APC and the PC state with respect to the APC state ($\Delta E_{APC-PC}$) being $23.565$ meV. Furthermore, the SCDM of both APC and PC states shows that no magnetic moment is induced on the Gr layer. This implies that the localized state of the vacancy in hBN(V$_B$) does not give a magnetic proximity effect on graphene. 

The LDOS of hBN(V$_B$)/Gr/hBN(V$_B$) in figure \ref{fig:figure4}(c) and (d) for the APC and PC states, respectively, shows that no proximity effect is also found in the LDOS of the Gr layer. This was shown from the localized state of hBN(V$_B$) from $E-E_F =0.5$ eV to $1.8$ eV in the spin minority channel, which does not affect the graphene LDOS in the same energy range. However, from the LDOS of Gr, the Gr layer gave a small magnetic response due to the magnetic configuration of the hBN(V$_B$) layers. In the case of APC, the Gr LDOS of the spin majority and minority channels is the same, indicating no magnetic response on the Gr layer. However, in the PC state, a small Stoner gap was observed, indicating the magnetic response of the Gr layer due to the magnetization of hBN(V$_B$) from both the upper and lower sides. 
Furthermore, the small mass-gapped Dirac cone of graphene appeared in both APC and PC states, which is consistent with a previous study \cite{quhe_2012}. This is due to the fact that the B atoms of hBN(V$_B$) were located on top and below one of the C atoms of graphene, slightly modulating the potential between the C atoms sublattice A and B of graphene. However, here a small shift of the mass-gapped Dirac cone of graphene was observed from zero energy to $0.5$ eV, which could come from the doping effect hBN(V$_B$).

\begin{figure}[tb]
\includegraphics[width=\columnwidth]{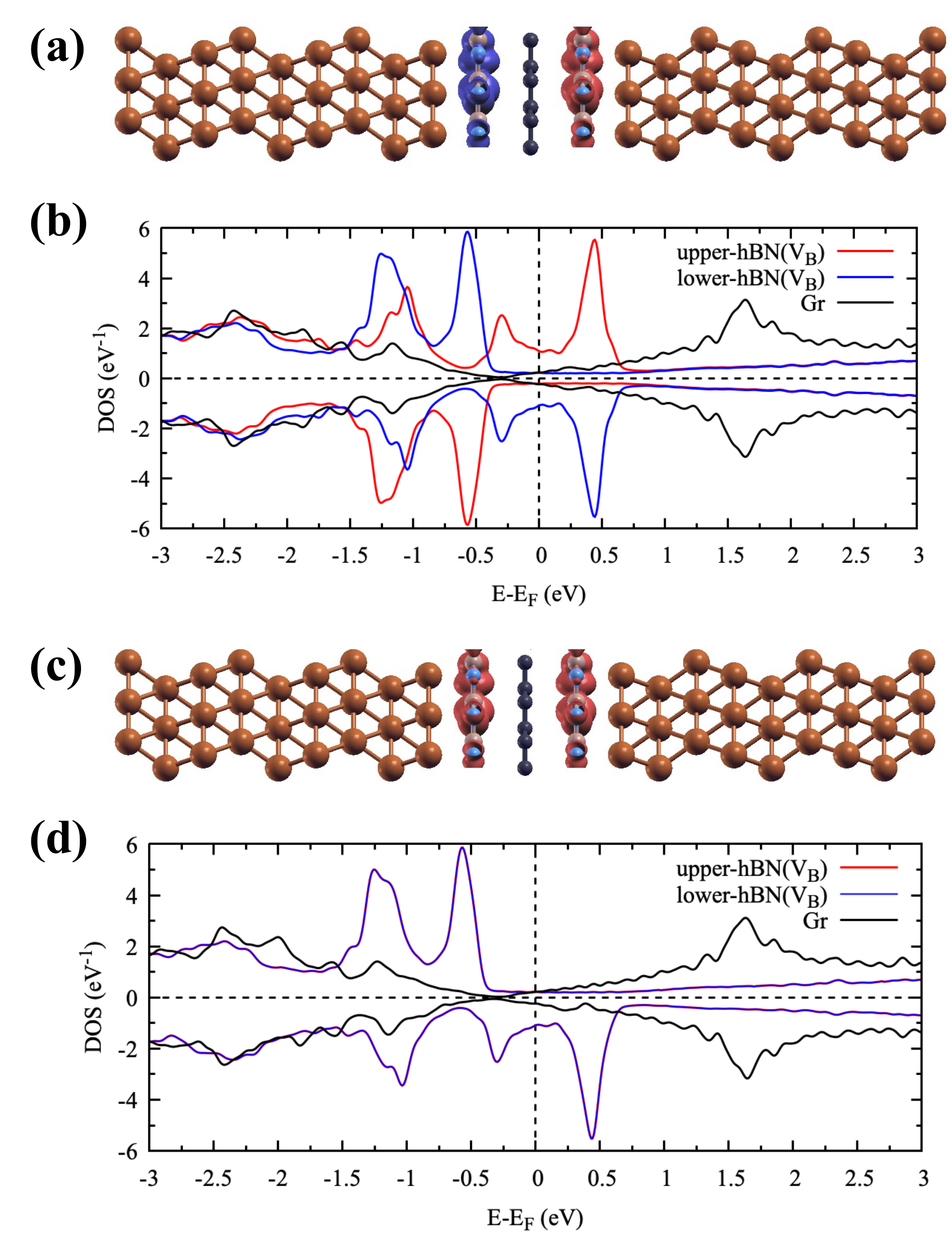}
\caption{\label{fig:figure5} SCDM of $3 \times 3$ Cu/hBN(V$_B$)/Gr/hBN(V$_B$)/Cu for (a) APC and (c) PC states (red (blue) color represents spin-up (spin-down) electorn density). LDOS of $3 \times 3$ Cu/hBN(V$_B$)/Gr/hBN(V$_B$)/Cu for (b) APC and (d) PC states. The positive (negative) value of DOS represents the spin majority (minority) channel.}
\end{figure}

\begin{figure}[tb]
\includegraphics[width=\columnwidth]{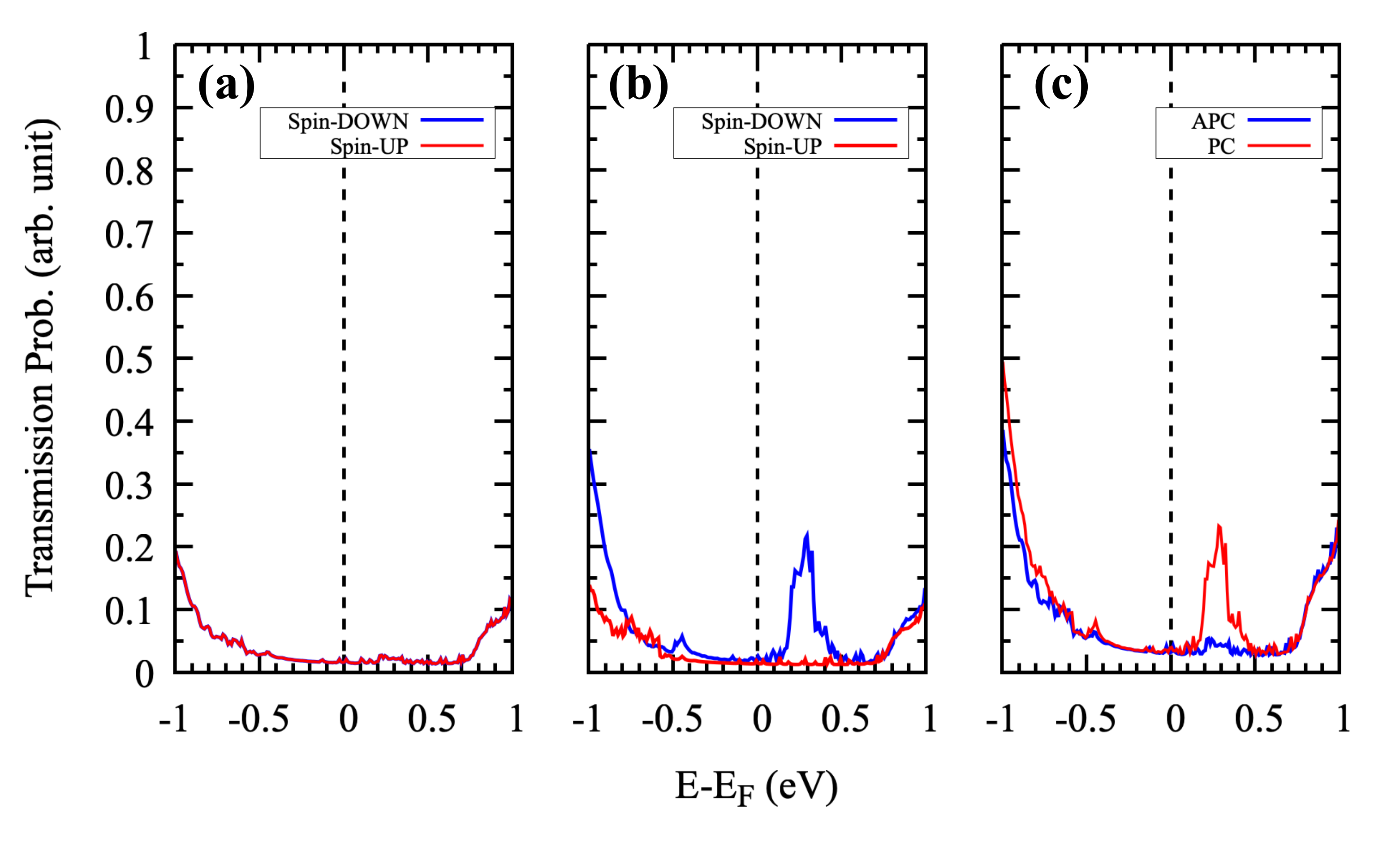}
\caption{\label{fig:figure6} Tunneling transmission probability of Cu/hBN(V$_B$)/Gr/hBN(V$_B$)/Cu for (a) APC and (b) PC states, and (c) the comparison between both of them.}
\end{figure}

The perfect spin-valve LDOS from $E-E_F=0.5$ eV to $1.8$ eV in hBN(V$_B$)/Gr/hBN(V$_B$) are expected to give a high efficiency of the TMR ratio. By connecting the hBN(V$_B$)/Gr/hBN(V$_B$) MTJ with the Cu electrode, the CPP transmission probability of Cu/hBN(V$_B$)/Gr/hBN(V$_B$)/Cu was investigated. Investigation was done first at Cu/hBN(V$_B$)/Gr/hBN(V$_B$)/Cu in APC state. The SCDM of the system shown in Figure \ref{fig:figure5}(a) suggests that there is no induced magnetic moment on the Cu layer at the interface, the same as in the case of Cu/hBN(V$_B$)/Cu. The LDOS of hBN(V$_B$) and Gr layers in Cu/hBN(V$_B$)/Gr/hBN(V$_B$)/Cu was shown in Figure \ref{fig:figure5}(b). Unlike the LDOS of Gr in hBN(V$_B$)/Gr/hBN(V$_B$) where small mass-gapped Dirac cone was observed, in Cu/hBN(V$_B$)/Gr/hBN(V$_B$)/Cu Gr layer keep its Dirac cone. This is due to the difference in the distance between layers between hBN (V$_B$) and the Gr layer in hBN(V$_B$)/Gr/hBN(V$_B$) and Cu/hBN(V$_B$)/Gr/hBN(V$_B$)/Cu which $2.83$ \AA~ and $2.88$ \AA, respectively. Furthermore, it was shown that the Fermi energy in Cu/hBN(V$_B$)/Gr/hBN(V$_B$)/Cu is shifted to the higher energy compare to hBN(V$_B$)/Gr/hBN(V$_B$). This shift could come from the surface state of the Cu electrode, giving the electron-doped effect to hBN(V$_B$)/Gr/hBN(V$_B$) (See Supplementary Materials). Although there is small additional LDOS due to the proximity effect of Cu on hBN(V$_B$), the opposite high intensity between the majority of spin and the minority channel near the vicinity of Fermi energy is still prominent and is expected to give electrons a low transmission probability. This low transmission probability was confirmed in Figure \ref{fig:figure6}(a). 

When considering the PC state, similarly, the SCDM of the system shown in Figure \ref{fig:figure5}(c) suggests that there is no induced magnetic moment in the Cu layer at the interface. From the LDOS, it also shows the absence of a graphene mass-gapped Dirac cone due to the increase of the interlayer distance between hBN(V$_B$) and the Gr layer and the shift of Fermi energy due to the electron-doped effect of the surface state of the Cu electrode. However, in this PC state, the upper and lower hBN(V$_B$) spin majority and minority channels have the same LDOS, creating a high possible transmission of electrons, especially at the DOS peak near $0.4$ eV. This high transmission probability around $0.4$ eV was confirmed in Figure \ref{fig:figure6}(b). Finally, when comparing the total transmission probability of APC with the PC state, a high TMR ratio $\approx 400\%$ was shown in Figure \ref{fig:figure6}(c). This high transmission probability was achieved by only considering the three-atomic thickness of MTJ based on 2D materials. 

The interchange between the PC and the APC states can be considered utilizing the small energy difference of $23.565$ meV. The PC state was previously found to be the most stable magnetic alignment between the upper and lower hBN(V$_B$) layers. Thus, to consider the APC state, the magnetic moment of one of the hBN(V$_B$) layers needs to flip to have an opposite direction. One of the methods that can be used is the spin accumulation effect on the Cu surface through spin vorticity coupling of the oxidized Cu \cite{Tatara, Nozaki}. When an in-plane current is given to oxidized Cu in a certain direction, spin accumulates on the surface of Cu near hBN(V$_B$), giving a magnetic field in the opposite direction of another hBN(V$_B$) layer. To change the magnetic moment configuration back to PC, an opposite in-plane current is introduced on oxidized Cu, creating the same spin direction accumulated on the surface of Cu near hBN(V$_B$). In addition, the small energy difference between the APC and PC states indicates that a small current was needed to change the magnetic configuration, resulting in a low-energy-consumption MTJ device.

This hBN(V$_B$)/Gr/hBN(V$_B$) can be fabricated using various methods; one of which involves employing chemical vapor deposition (CVD) to create a stack layer of hBN / Gr / hBN \cite{CVD1,CVD2,CVD3} and introducing a monoatomic vacancy on hBN using Ar ion sputtering \cite{Ar-ion}. Initially, hBN CVD growth on the clean surface of Cu(111) substrate can be achieved by introducing common Boron and Nitrogen precursors such as Boron trichloride (BCl$_3$) or Boron trifluoride (BF$_3$) and Ammonia (NH$_3$), respectively. Subsequently, a monoatomic vacancy was introduced on the pristine hBN layer by using Ar-ion sputtering. Following this, the Gr layer was growth on hBN by introducing precursor gases of carbon atoms, such as methane (CH$_4$) or ethylene (C$_2$H$_4$). And finally, the second hBN(V$_B$) is grown in the same manner as the first hBN(V$_B$).

\section{Conclusion}
In this work, we proposed a novel three-atom-layer van der Waals-based MTJ utilizing boron-vacancy defects in hBN. The approach involves employing both DFT for calculating electronic and magnetic properties and the LB formalism within the NEGF method to determine the MTJ's transmission probability. Our findings observed the formation of a Stoner gap between the spin-majority and spin-minority channels on the LDOS of hBN(V$_B$) near the Fermi energy. This enables control over the spin-valve behavior by manipulating the magnetic alignment (APC and PC) of the hBN(V$_B$)layers. The results showed that the PC state exhibits high electron transmission, whereas the APC state results in low electron transmission, resulting in a remarkable TMR ratio of approximately 400\%. These findings suggest promising prospects for developing high-performance spintronics devices with the proposed three-atom-layer MTJ design.


\section*{Acknowledgements}
Calculations were performed at the Kyushu University computer center. H.H. gratefully acknowledges the fellowship support from the JSPS. This study was partly supported by the Japan Society for the Promotion of Science (JSPS) KAKENHI (Grant Nos. JP19H00862 and JP16H00914 in the Science of Atomic Layers, 21J22520 in the Grant-in-Aid for Young Scientists, and JP18K03456).

\bibliography{apssamp}

\end{document}